\documentclass[prl,twocolumn]{revtex4}
\usepackage{amsfonts}
\usepackage{amsmath}
\usepackage{amssymb}
\usepackage{graphicx}

\newcommand{\BE}{\begin{equation}}
\newcommand{\BEA}{\begin{eqnarray}}
\newcommand{\EE}{\end{equation}}
\newcommand{\EEA}{\end{eqnarray}}

\newcommand{\om}{\omega}

\begin{document}

\title{Competition between Antiferromagnetism and Superconductivity in High $T_c$ Cuprates.}
\author{David S\'{e}n\'{e}chal, P.-L. Lavertu, M.-A. Marois, and A.-M.S. Tremblay}
\affiliation{D\'{e}partement de physique and Regroupement qu\'{e}b\'{e}cois sur les mat\'{e}riaux de pointe, Universit\'{e} de Sherbrooke, Sherbrooke, Qu\'{e}bec, Canada, J1K 2R1}
\date{October 2004}

\begin{abstract}
Using variational cluster perturbation theory we study the competition between d-wave superconductivity (dSC) and antiferromagnetism (AF) in the the $t$-$t'$-$t''$-$U$ Hubbard model.
Large scale computer calculations reproduce the overall ground state phase diagram of the high-temperature superconductors as well as the one-particle excitation spectra for both hole- and electron-doping. We identify clear signatures of the Mott gap as well as of AF and of dSC that should be observable in photoemission experiments.
\end{abstract}

\maketitle


Ever since the first paper of P.W. Anderson on high-temperature superconductivity \cite{Anderson:1987}, most theorists have been working with the premise that the physics of these most intriguing materials should be contained in the two-dimensional one-band Hubbard model.
That model of strongly interacting electrons on a two-dimensional square lattice contains a kinetic energy term that represents the band structure, and a potential energy term that accounts only for on-site (perfectly screened) interaction.
This simplicity may be too naive, in particular in light of the striking contrast between the phase diagram of the electron-doped and that of the hole-doped cuprates.
The former has an antiferromagnetic (AF) phase that extends over a wide range of doping while the d-wave superconducting (dSC) phase seems restricted to a small range.
The reverse is true for the hole-doped case.
In addition, some hole-doped cuprates seem to exhibit stripes \cite{arrigoni:2004} or other recently discovered phases \cite{lupien:2004} near the boundary between antiferromagnetism and d-wave superconductivity.
We cannot trust that copper-oxygen planes described by the Hubbard model contain the whole physics until we have shown that both the hole- and electron-doped cuprates can be explained by this model.
Yet previous theoretical studies that found agreement with experiment  \cite{Randeria:2004} concentrated on the hole-doped cuprates.
In addition, many authors argue that the one-band Hubbard model description of the cuprates is inadequate to obtain dSC \cite{Zhang97, Jarrell:2004} or that additional physical effects must be invoked \cite{Assaad:1996, Lanzara:2004}.

In this letter we show, in agreement with the basic premise of Anderson, that the two-dimensional Hubbard model accounts for the observed doping dependence of the AF and dSC phases as well as for the single-particle excitations for both electron and hole doping.
The striking asymmetry between electron and hole doping can be accounted for by the fact that the known band structure is not particle-hole symmetric and that interactions at intermediate-coupling cannot completely wipe out band effects.
We also identify distinct spectral features of Mott, AF and dSC physics.

Our results were obtained thanks to a new methodology, namely Variational Cluster-Perturbation Theory (V-CPT) \cite{Potthoff03}, and because of remarkable increases in computer power made possible by cluster architectures.
We study the Hubbard model with interaction $U$ and with
hopping parameters that are taken from band structure calculations \cite{Andersen95}, namely a diagonal hopping $t'=-0.3t$ and third-neighbor hopping $t''=0.2t$.


\paragraph{Variational cluster perturbation theory.}

V-CPT is an extension of Cluster Perturbation Theory (CPT) \cite{Senechal00} that is based on the self-energy-functional approach (SFA) \cite{Potthoff:2003}.
This approach uses the rigorous variational principle $\delta \Omega _{\mathbf{t}%
}[\Sigma ]/\delta \Sigma =0$ for the thermodynamic grand-potential $\Omega _{%
\mathbf{t}}$ written as a functional of the self-energy $\Sigma $
\begin{equation}
\Omega _{\mathbf{t}}[\Sigma ]=F[\Sigma ]+\mathrm{Tr}\ln
(-(G_0^{-1}-\Sigma )^{-1}).
\end{equation}%
The index $\mathbf{t}$ denotes the explicit dependence of $\Omega _{\mathbf{t}}$ on the matrix $t_{ij}$ of hopping terms or, more generally, on {\it all one-body operators}. That dependence comes through the Green function $G_0$ of the one-body part of the Hamiltonian.
In the above expression, $F[\Sigma ]$ is a universal functional of the self-energy obtained from the Legendre transform of the Luttinger-Ward functional.
The physical Green function is $G=-\delta F/\delta \Sigma $ \cite{Potthoff03} and the stationary condition for $\Omega
_{\mathbf{t}}[\Sigma ]$ give Dyson's equation.
Although $F$ is universal, its exact form is unknown.
But all Hamiltonians with the same interacting part share the same functional form of $F[\Sigma ]$.
Hence $F[\Sigma ]$ may be derived from the exact solution of a simpler Hamiltonian $H'$ whose choice of one-body terms makes it exactly solvable. One then looks for stationary solutions within the subspace defined by that simpler solvable problem.
The Dynamical Mean-Field theory (DMFT) \cite{DMFT} is a special case \cite{Potthoff03} where the simpler problem $(H')$ is that of a single site (the
``impurity'') connected to an infinite bath of non-interacting sites.
When the simpler problem consists of a single cluster connected to a bath, one recovers Cellular-DMFT \cite{Kotliar01}. The Dynamic Cluster Approximation (DCA) \cite{Jarrell:2004, DCA, Maier04} does not fit in that general variational scheme \cite{Potthoff03}, although it is self-consistent.
In V-CPT, one uses for $H'$ a Hamiltonian formed of clusters that are disconnected by removing hopping terms between identical clusters that tile the infinite lattice.
In the cluster, the Hamiltonian contains the original Hubbard interaction, hopping, and Weiss fields that are going to be determined by minimizing the grand potential $\Omega _{\mathbf{t}}$.
In V-CPT the cluster shape and size can be varied as a check of the validity of the results in the thermodynamic limit.
V-CPT is thermodynamically consistent and causal. It goes beyond ordinary mean-field theory, since it predicts the absence of long-range AF order in one dimension \cite{Dahnken03}.

In practice, one proceeds as follows.
The universality of $F$ allows us to express the functional $\Omega _{\mathbf{t}}$ in terms of the corresponding functional for the cluster Hamiltonian $H'$, i.e., $\Omega _{%
\mathbf{t}'}$ in Eq.(5) of Ref.~\cite{Potthoff:2003}.
To include the possibility of broken symmetry, $H'$ contains the terms
\begin{eqnarray}
H_{M}&=&M\sum_{a}(-1)^{a}(n_{a\uparrow }-n_{a\downarrow }) \\
H_{D}&=&\sum_{a,b}\Delta _{ab}\left( c_{a\uparrow }c_{b\downarrow }+\mathrm{%
H.c.}\right)
\end{eqnarray}
where $\Delta _{ab}=D$ if sites $a$ and $b$ are nearest-neighbors (NN) along the $x$ axis and $\Delta _{ab}=-D$ if the sites are NN along the $y$ axis. Such a Weiss field cannot be obtained from a Hartree-Fock factorization of the Hubbard model.
The Weiss fields $M$ for AF and $D$ for dSC phases allow for the physics of long-range order while the dynamical short-range interactions are taken into account in the exact diagonalization of the cluster.
One then restricts the space of self-energies to the exact self-energies of all possible cluster Hamiltonians: $\Sigma =\Sigma (\mathbf{t}')$.
In that case the set of variables $\mathbf{t}'$ includes the variational parameters $M$ and $D$.
Rearranging Eq.(5) of Ref.~\cite{Potthoff:2003} in terms of the cluster grand-potential, $\Omega'\equiv \Omega'_\mathbf{t}[\Sigma]$ and Green function
$G'{}^{-1}\equiv G'_0{}^{-1}-\Sigma$, one finds
\begin{equation}
\Omega _{\mathbf{t}}(\mathbf{t}')=\Omega'\kern-0.4em - \kern-0.4em\int_C{\frac{%
d\omega }{2\pi }}\sum_{\mathbf{K}}\ln \det \left(
1+(G_0^{-1}\kern-0.2em -G_0'{}^{-1})G'\right),
\end{equation}%
which is the starting point of numerical calculations.
The functional trace has now become an integral over the diagonal variables (frequency and superlattice wave vectors) of the logarithm of a determinant over intra-cluster indices.
The frequency integral is carried along the imaginary axis, since it can be shown that the integrand decreases asymptotically as $1/(i\omega )^{2}$.
The minimum of $\Omega(M,D)$ is found using the conjugate-gradient algorithm.
Computing time varies a lot, from one cpu-hour in the case of a 4-site cluster, to hundreds of cpu-hours for a 10-site cluster.


\paragraph{Order parameters.}

The treatment of superconductivity in V-CPT is best done in the Nambu representation.
Translation invariance in the superlattice implies that the Green function $G{}^{-1}\equiv G_0{}^{-1}-\Sigma$ depends, after Fourier transformation, on two wave vectors $\mathbf{k}$ and $\mathbf{k}'$ that can only differ by a reciprocal superlattice vector.
The electron concentration ($n$), the AF order parameter ($%
M_{o}$) and the dSC order parameter ($D_{o}$) are respectively expressed as
\begin{eqnarray}
\label{OPeq}
n &=&2i\int {\frac{dk}{(2\pi )^{2}}}\int_{C}{\frac{d\omega}{2\pi}}\sum_\sigma%
\mathcal{G}_\sigma(\mathbf{k},\mathbf{k},\omega)   \\
M_{o} &=&2i\int {\frac{dk}{(2\pi )^{2}}}\int_{C}{\frac{d\omega }{2\pi }}%
\sum_\sigma (-1)^\sigma \mathcal{G}_\sigma(\mathbf{k},\mathbf{k}+\mathbf{Q},\omega ) \\
D_{o} &=&2i\int {\frac{dk}{(2\pi )^{2}}}\int_{C}{\frac{d\omega }{2\pi }}%
\mathcal{F}(\mathbf{k},\mathbf{k},\omega )g(\mathbf{k})
\end{eqnarray}%
where $\mathbf{Q}=(\pi ,\pi )$ is the antiferromagnetic wave vector, $\mathcal{G}_\sigma$ the normal and $\mathcal{F}$ the anomalous V-CPT Green functions, written now as functions of wavevector. The dSC form factor is $g(\mathbf{k})=\cos(k_x)-\cos(k_y)$.
The integral over frequencies is carried along the upper half of a clockwise contour that encloses the poles of the Green function up to $\omega=0$.



\paragraph{Results.}

We present V-CPT calculations on $L=$ 6-site ($2\times 3$), 8-site ($2\times 4$) clusters and the 10-site cluster of Ref.~\cite{Dahnken03}, using $M$ and $D$ as simultaneous variational parameters and $U$ and $\mu$ (the chemical potential) as control parameters.
The number of electrons within the cluster ($N_{c}$) is a conserved quantity only when the dSC Weiss field $D$ vanishes.
When $D\neq 0$, $n_{c}$ ($n_{c}=N_{c}/L$) can take any real value controlled by $\mu$.
The calculated electron concentration $n$ (Eq.~\ref{OPeq}) is not quantized, even when $D=0$, except when $\mu$ lies within the Mott gap ($n=n_{c}=1$).
Even though results are displayed as functions of $n$, one must keep in mind that it is the chemical potential $\mu$ that is the true control parameter.

\begin{figure}[tbp]
\centerline{\includegraphics[width=7.5cm]{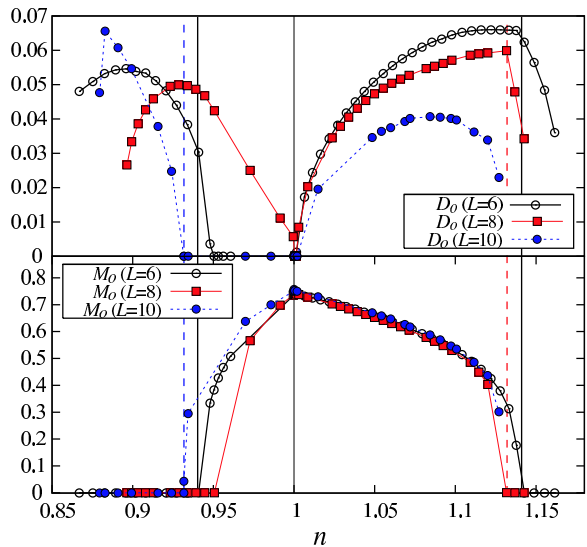}}
\caption{AF (bottom) and dSC (top) order parameters for $U=8t$ as a function of the electron density ($n$) for $2\times3$, $2\times4$ and 10-site clusters. Vertical lines indicate the first doping where only dSC order is non-vanishing.}
\label{fig_8Un}
\end{figure}

\begin{figure}[tbp]
\centerline {\includegraphics[width=7.5cm]{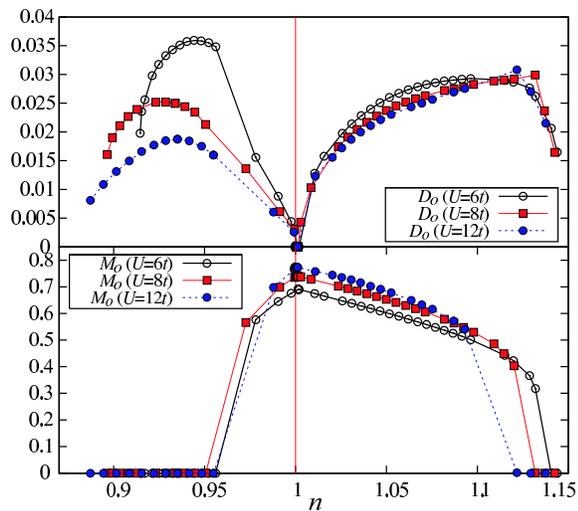}}
\caption{AF and dSC order parameters as a function of the electron density on a $2\times 4$ cluster for $U=6t$, $U=8t$ and $U=12t$.}
\label{fig_UMD}
\end{figure}

The Weiss fields $M$ and $D$ as a function of $\mu$ are obtained from the stationary point (in this case a minimum) of the grand potential.
The Weiss fields generally decrease with increasing cluster size, as it should for spontaneous symmetry breaking. 
Instead of the Weiss fields, we show the order parameters as a function of the electron concentration $n$ calculated from Eq.~\ref{OPeq}, for different system sizes in  Fig.~\ref{fig_8Un} and for different interaction strengths in Fig.~\ref{fig_UMD}.

The bottom panel showing $M_0$ in Fig.~\ref{fig_8Un}
clearly reproduces the striking contrast between electron- $\left(
n>1\right) $ and hole- $\left( n<1\right) $ doped systems.
Antiferromagnetism can extend up to about $15\%$ doping on the electron-doped side while it persists up to only about $6\%$ on the hole-doped side. The dependence on system size is not monotonic because even at fixed cluster size there can be a dependence on the shape of the cluster.
Nevertheless, there is a clear convergence with system size, especially on the electron-doped side.  
The dSC order parameter is shown on the top panel. For different system sizes, the vertical lines indicate the filling where the dSC phase appears by itself, without AF order parameter. 
In the electron-doped case, the pure dSC phase is on the right. It survives for a narrow range only, namely $n=1.13$ to $n=1.15$ for $L=8$ for example. 
In the hole-doped case, the dSC phase on the left of the vertical lines is present for a range of fillings that is at least twice as large, $n=0.87$ to $n=0.93$ for $L=10$ for example. Unfortunately, finite cluster effects do not allow us to obtain reliable results for larger dopings for either the hole or electron cases. The dSC order parameter is not present on bonds between clusters, making the results more sensitive to size effects than in the case of the AF phase.
Fig.~\ref{fig_8Un} also shows, in the electron-doped case, a AF+dSC phase where AF and dSC order parameters are both non-vanishing \cite{Inui:1988, Lichtenstein:2000, Sachdev:2001}.
We verified that, as expected from symmetry, the $\pi$-triplet order parameter is non-vanishing in that phase\cite{Fenton:1983, Kyung:2000}, which is separated from a pure dSC phase by a quantum critical point around 13\% doping, near the value suggested by experiment \cite{Greene:2004, Fournier:1998}.
The situation is less clear on the hole-doped side where the $L=6$ cluster has a very small doping range for the AF+dSC phase, the $L=8$ cluster a large one, while the $L=10$ cluster shows none. This suggests that the way in which the AF and dSC phases approach each other on the hole-doped side cannot be accurately described by the small variational space that we use. Additional order parameters, such as stripe \cite{arrigoni:2004} or checkerboard orders \cite{lupien:2004} observed in certain cuprates may be necessary to get the full picture.  No $SO(5)$ symmetric point \cite{Zhang:1997} appears in our calculation in a size-independent way. On the other hand, our results for $D_0$ in Fig.~\ref{fig_8Un} show unambiguously that the pure dSC phase appears over a much broader range of dopings for hole- than for electron-doped cuprates, as observed experimentally.

It is also instructive to know how the ground-state order parameters vary with interaction strength $U$, especially because several normal-state calculations for the pseudogap \cite{Senechal04, Kyung:2004} show that the interaction strength for electron-doped cuprates near optimal doping should be in the weak to intermediate coupling range $\left( U\sim 6t\right)$, with $U$ increasing as $n$ decreases.  A look at Fig.~\ref{fig_UMD} for $D_0$ and $M_0$ shows that the range of dopings where only $D_0$ is non-vanishing is larger on the hole than on the electron-doped side for all values of $U$.  That range increases with $U$ in all cases so that a drop in $U$ as $n$ increases reinforces the electron-hole difference in the size of the pure dSC region.
The range where only the dSC order parameter is finite nearly doubles in going from $U=6t$ to $U=12t$ but the maximum value of $D_0$ decreases, at least on the hole-doped side.
A stronger repulsion allows systems with more holes to be superconducting, but at the same time suppresses superconductivity more effectively closer to half-filling.
Note that the dSC order parameter $D_0$ should not be confused with the critical temperature: The maximum in $D_0$ that appears near $n=0.9$ on the hole-doped side does not mean that the maximum $T_c$ is around that doping.
Instead, the maximum comes from the growth of $D_0$ towards half-filling until proximity with the AFM phase makes it fall rapidly \cite{Kyung:2000}. Finally, as seen in previous calculations, as $U$ increases from $6t$ to $12t$, the AF phase does not extend as far on the electron-doped side, \cite{Freericks:1994} whereas the value of the order parameter $M_0$ at half-filling increases \cite{Fano:1992}.

\begin{figure}[tbp]
\centerline {\includegraphics[width=8cm]{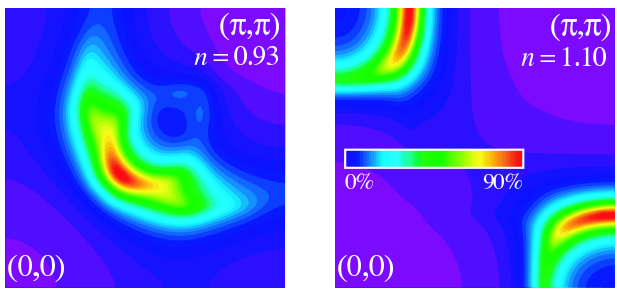}}
\caption{Intensity plot of the spectral function at the Fermi level, in the first quadrant of the Brillouin zone, for $U=8t$ on a $L=8$ cluster.
Left: Hole-doped system ($n=0.93$).
Right: Electron-doped systems ($n=1.10$). A Lorentzian broadening of $0.2t$ is used.}
\label{fig_mdc}
\end{figure}

\begin{figure}[tbp]
\centerline {\includegraphics[width=8cm]{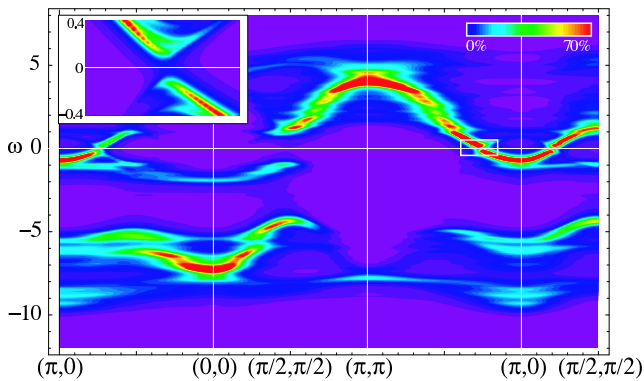}}
\caption{Intensity plot of the spectral function as a function of $\protect%
\omega$ in units of $t$, and wave vector, for the same parameters as the right-hand panel of Fig.~3. The Lorentzian broadening is $\protect\eta=0.12t$ in the main figure and $\protect\eta=0.04t$ in the inset.}
\label{fig_spectre}
\end{figure}

Fig.~\ref{fig_mdc} shows intensity plots of the spectral functions at the Fermi level, for $U=8t$, in the first quadrant of the Brillouin zone.  The left illustrates a hole-doped system in a pure dSC phase. The spectral weight is concentrated along the diagonal.
This is observed even without long-range order \cite{Senechal04}, but is also compatible with the vanishing of the dSC gap along the diagonal. On the right, we display an electron-doped system in a AF+dSC phase.
The spectral weight is depleted along the diagonal and concentrated near the zone boundaries ($(\pi ,0)$ and $(0,\pi )$).  This is also observed in the absence of long-range order \cite{Senechal04}, but long-range order makes the plots sharper.  The vanishing of the usual dSC gap along the diagonal is offset by the AF gap.
The above asymmetric behavior of the spectra on electron- and hole-doped sides is observed in Angle Resolved Photoemission Spectroscopy (ARPES) \cite{Armitage02, Ronning03}.

The Mott phenomenon, as well as the dSC and AF order parameters, each have distinct signatures in the spectral function. 
Fig.~\ref{fig_spectre} illustrates this in the AF+dSC phase for the electron-doped case. 
The Mott gap, already seen in ARPES \cite{Armitage02}, corresponds to an absence of states for all wave vectors in the interval $-4t<\omega<-2t$. 
The two main bands that disperse in the intervals $-8t<\omega<-4t$ and $-t<\omega<5t$ roughly correspond to those obtained in AF mean-field theory~\cite{Kusko02}.
The AF order parameter manifests itself through the ``shadowing'' of these main bands by reflection about the magnetic zone boundary.
The broad, dispersionless features around $\om\sim-10t$ and $5t$ are remnants of the atomic limit, intimately related to the Mott phenomenon.
The excitations occurring around $\om\sim-2t$ are absent at half-filling and from the mean-field solution.
Their presence considerably reduces the direct gap at $(\pi/2,\pi/2)$.
Hints of these new excitations, together with the AF shadowing within the lower Hubbard band, are seen in Fig.3 of Ref.~\cite{Armitage02} for $n=1.04$.
The inset, which is a higher-resolution blowup of the rectangular region shown in the figure on the $(\pi,\pi)$ to $(\pi,0)$ segment, shows the dSC gap.



\paragraph{Conclusion.}
This work shows, within a rigorous variational approach that takes into account short-range dynamics using clusters of different sizes, that the one-band Hubbard model contains the essential physics of the cuprates. It has dSC and AF ground states in doping ranges that are consistent with the observed ones for both electron- and hole-doping. In addition, low energy excitations manifest in the electronic spectra are shown to be strongly momentum dependent in essentially the same manner as observed using ARPES. Our results also suggest that a quantum critical point on the electron-doped side separates a pure dSC phase from a phase where both AF and dSC order parameters are non-vanishing. 
In the latter phase, we identified distinct spectral features associated with the Mott gap and with AF and dSC long-range orders that should be observable by ARPES.


\begin{acknowledgments}
We are indebted to V. Hankevych, S. Kancharla, B. Kyung, and M. Norman for useful discussions.
This work was supported by NSERC (Canada), FQRNT (Qu\'ebec), CFI (Canada), CIAR and the Tier I Canada Research Chair Program (A.-M.S.T.).
Computations were performed on a 200-cpu Beowulf cluster and on the 872-cpu Dell cluster of the RQCHP.
\end{acknowledgments}



\end{document}